\documentclass[twocolumn, secnumarabic, amssymb, superscriptaddress, aps, prl]{revtex4-1}
\usepackage{amsmath,graphicx,float,color}
\usepackage{bm}
\usepackage{amssymb}
\usepackage{colordvi}
\usepackage{graphicx}
\usepackage{color}
\usepackage{hyperref}
\usepackage{ulem}
\usepackage{color}

\begin{document}
\bibliographystyle{unsrt}

\title{Revisiting the validity of the Stokes-Einstein relation}


\author{Hanqing Zhao}
\email{Current affiliation: Department of Physics, University of Colorado Boulder}
\affiliation{Department of Physics, Xiamen University, Xiamen 361005, Fujian, China}
\affiliation{Department of Modern Physics, University of Science and Technology of China, Hefei 230026, China}
\author{Hong Zhao}
\email{zhaoh@xmu.edu.cn}
\affiliation{Department of Physics, Xiamen University, Xiamen 361005, Fujian, China}


\begin{abstract}
The Stokes-Einstein (SE) relation has been widely applied to quantitatively describe the Brownian motion. Notwithstanding, here we show that even for a simple fluid, the SE relation may not be completely applicable. Namely, although the SE relation could be a good approximation for a large enough Brownian particle, we find that it induces significant error for a smaller Brownian particle, and the error increases with the decrease of the Brownian particle's size, till the SE relation fails completely when the size of Brownian particle is comparable with that of a fluid molecule. The cause is rooted in the fact that the kinetic and the hydrodynamic effects depend on the size of the Brownian particle differently. By excluding the kinetic contribution to the diffusion coefficient, we propose a revised Stokes-Einstein relation and show that it expands significantly the applicable range.

\end{abstract}
\maketitle

The Stokes-Einstein (SE) relation,
\begin{equation}
D R=\frac{k_BT}{c\eta},
\end{equation}
establishes a connection between the diffusion coefficient, $D$, of a Brownian particle and the shear viscosity, $\eta$, of the fluid it is immersed in. Here $R$ is the radius of the Browinan particle, $k_B$ is the Boltzmann constant, $T$ is the fluid temperature, and $c$ is a constant depending on the boundary conditions. It makes the Einstein relation~\cite{einstein},  $\langle r(t)^2\rangle=2Dt$, a quantitative law. Although the SE relation has been found invalid in certain extreme situations, such as in supercooled liquids\cite{rossler,dzugutov2002decoupling}, in glass-forming liquids~\cite{becker2006fractional,mazza2007connection,xu2009appearance,dehaoui2015viscosity}, and in dense complex medias~\cite{rosenfeld1999quasi,dzugutov2001addendum,bretonnet2002self,kaur2005nature,ning2019universal}, it has been a long-standing and well acknowledged belief that for a simple fluid the SE relation applies well even when the size of the Brownian particle decreases down to the molecular level~\cite{mason1997particle,barhoum2016diffusion,palit2017combining,stillinger1994translation,wilbur1976self,parkhurst1975dense,harris2016scaling,harris2009fractional,alder1970,tang1995shear,ohtori2018stokes,liu2006test}.
Consequently, this law is routinely used to compute $D$ via the measurement of $\eta$, or to estimate the radius of molecules with measured $D$ and $\eta$.

However, a comprehensive testing of the SE relation is still insufficient though over a century has passed. Previous verifications, experimentally~\cite{rossler, parkhurst1975dense,wilbur1976self,stillinger1994translation,dzugutov2002decoupling, becker2006fractional,mazza2007connection,dehaoui2015viscosity} or  numerically~\cite{alder1970,parkhurst1975dense,tang1995shear,liu2006test,harris2009fractional,harris2016scaling,ohtori2018stokes}, have focused on the special case where a tagged fluid molecule is adopted as the Brownian particle. Moreover, what these studies have checked is only the linear relationship between $D$ and $T$ of the SE relation with a fixed $R$. Whether and under what conditions $DR$ is independent of the size $R$ and mass $M$ of the Brownian particle for a given fluid have not been systematically studied, to our knowledge. Note that the size and mass dependence are of importance for applications, since in practice the radiuses of a Brownian particle can range from nanometer to millimeter ~\cite{fouchier2012restricted,shindell2019climate} and the mass may vary in a wide range as well.

In this paper, we study the size and mass dependence of the SE relation in the hard-sphere fluid model. A key step is to decompose $D$ into the kinetic and the hydrodynamic part, i.e. $D_K$ and $D_H$. In doing so, we find that $D_K\sim R^{-2}$ but $D_H \sim R^{-1}$. The scaling behavior of $D_K$ indicates that the product $DR$ does not remain $R$-independent when $D_K$ cannot be neglected, which provides a key evidence that the SE relation is not an accurate law. On the other hand, the scaling behavior of $D_H$ inspires us to modify the SE relation as

\begin{equation}
D_HR=\frac{k_BT}{c\eta}.
\end{equation}

We perform large-scale numerical simulations to examine Eqs. (1) and (2). The results suggest that Eq. (1) is applicable for a Brownian particle over a few hundred times larger than the fluid molecules, while Eq. (2) is accurate for a Brownian particle down to only several times larger than the fluid molecules. These results are in clear contrast with the common belief that the SE relation is applicable at the molecular level.

Methodologically we emphasize the necessity of the decomposition in applying particle diffusion laws. The fact that Eq. (2) holds instead indicates that to apply hydrodynamic laws, it is appropriate to exclude contributions from kinetic effects. Similarly, laws established on the basis of pure kinetic effects should be applied after excluding the hydrodynamic contributions. To support this proposition, we examine the Chapman-Enskog theory of the diffusion coefficient. It is a kinetic theory that accounts for only the binary collisions mutually uncorrelated~\cite{sl,bretonnet2002self}. Excluding the hydrodynamic effect, we find that the theoretical prediction agrees perfectly with the simulated $D_K$ over the entire fluid regime.

{\bf Model and algorithm.} We employ the hard-sphere fluid model~\cite{alder1,alder2,sl} to perform our tasks. It consists of a sphere Brownian particle of mass $M$ and radius $R$, and $N$ fluid molecules. Each molecule has a unit mass $m=1$ and a sphere shape of radius $b$. The packing fraction of fluid with radius $b$ is given by $\phi=(4/3)\pi b^3n$, where the number  density $n$ of fluid molecules is fixed at $n=10^{-3}$ in our study. Note that with $R=b$ and $M=1$, the Brownian particle reduces to a fluid molecule. Initially, all the entities are placed evenly in a cubic box of side length $L$ with periodic boundary conditions, and assigned a random velocity sampled from the equilibrium velocity distributions corresponding to temperature $T=1$ ($k_B$ is set to be unity) with the restriction that the total momentum is zero. Next, the system is evolved for a sufficiently long time to ensure that it has fully relaxed to the equilibrium state, then the velocity autocorrelation function (VACF), defined as $C(t)=\langle {\bf v(t)}\cdot {\bf v(0)}\rangle/3$, is calculated, where ${\bf v(t)}$ is the velocity of the Brownian particle at time $t$. The collision between any two constituent entities are completely elastic.

To evolving the system numerically, an event-driven molecular dynamics algorithm is adopted ~\cite{alder1,alder2,marin1997event}. As it is challenging to compute the VACF of a Brownian particle, the maximum number of fluid molecules and the maximum radius of the Brownian particle are set to be $N = 27,000$ and  $R\le25$, respectively. For more details of the simulations, see the Supplemental Materials.

The hard-sphere model has an essential advantage for our purposes here, i.e., the radius of the Brownian particle is definitely defined. As a result, once $DR$ is shown to vary with radius $R$ or mass $M$ in a given fluid (thus $\eta$ and $T$ are fixed), the violation of the SE relation can be concluded. To this end, only the diffusion coefficient needs to be calculated. The viscosity $\eta$ of the fluid is required only when we want to determine the exact value of $c$.

\section*{Results}
{\bf Decomposition of the diffusion coefficient.} The diffusion coefficient can be calculated by the Green-Kubo formula based on the VACF:

\begin{equation}
D(t)=\int_0^t C(t') dt'.
\end{equation}
 Earlier studies have shown that the diffusion of a particle is governed by both kinetics and hydrodynamics~\cite{alder1,alder2,cohen2,ernst1,cohen1,resi,ernst2}. Consequently, the VACF can be decomposed as  $C(t) = C_K(t) + C_H(t)$.

The kinetic approach to the VACF is pioneered by Einstein~\cite{einstein}. The key insight is that the decay of the VACF is induced by random collisions with surrounding fluid molecules, by which the Brownian particle loses its initial momentum exponentially.  The collisions are considered to be uncorrelated, which leads to~\cite{sl}

\begin{equation}
C_K(t) = C(0) \exp(-\frac{C(0)}{D_K}t),
\end{equation}
where $C(0) ={k_BT}/M$.

It has been known since the 1960s that the VACF also includes a hydrodynamic contribution: The momentum transferred to the fluid can feedback to the Brownian particle through velocity vortex field~\cite{alder1,alder2,ernst1,ernst2,resi} or, equally, through ring collisions~\cite{cohen1,cohen2}. The feedbacked momentum gives rise to
 $C_H(t)$. This effect can also be approximately formulated by the extended Langevin equation by involving a memory factor in the collisions~\cite{zwanzig1,russel}. These studies have established that $C_H(t)$ has a power-law tail at the long-time limit~\cite{sl}, i.e.,

\begin{equation}
C_H(t)\sim [4\pi(D_K+\nu)t]^{-\frac{3}{2}},
\end{equation}
where $\nu=\eta/(mn)$ is the kinematic shear viscosity.

Our method for separating $C_K(t)$ from $C(t)$ is as following. In a short time, the momentum transferred from the Brownian particle to the fluid is little due to few collisions; moreover, this part of transferred momentum does not contribute to the VACF of the Brownian particle, because a collision ring has not formed. A collision ring forms only when the momentum of the Brownian particle transferred to the fluid molecules comes back at a later moment to the Brownian particle itself through a collision chain of fluid molecules. Therefore, before the collision ring forms, $C(t)$ is identical to $C_K(t)$. In practice, we fit $C_K(t)$ by a proper $D_K$ with Eq. (4) in a given short time interval $(0, t)$. The hydrodynamic effect is regarded to play no role if the fitting result does not change by decreasing $t$ further.

\begin{figure}[!htbp]
\centering
\includegraphics[width=\linewidth]{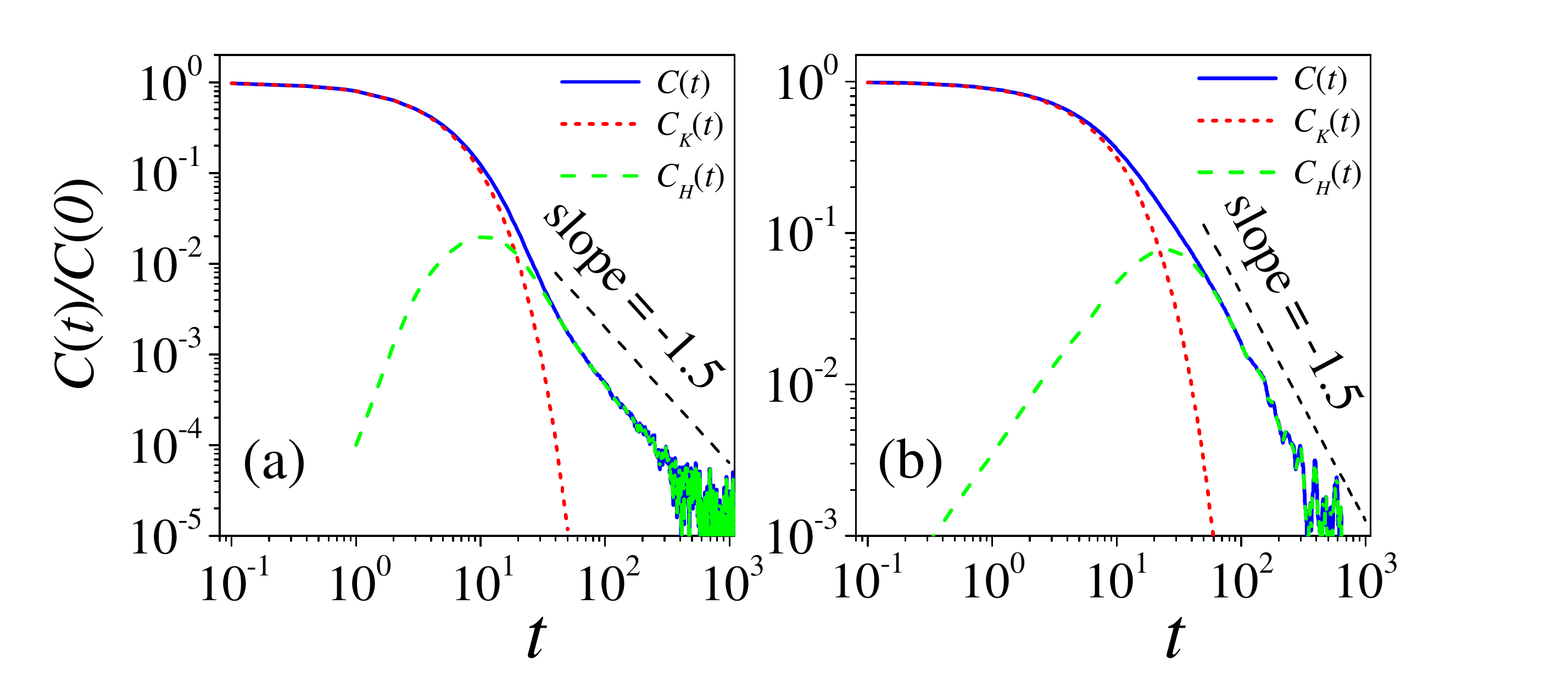}

{\caption{The VACF of the Brownian particle in the fluid of $\phi=0.11(b=3)$. (a) The Brownian particle is identical with the fluid molecules; $M=1$ and $R=b$; (b) $M=24$ and $R=20$.}}
\label{zhqfig1}
\end{figure}
Figure 1(a) and 1(b) show the VACF of the Brownian particle when it is identical with and different from the fluid molecules, respectively. With $D_K$ determined by the best fitting introduced above, we obtain $C_K$ and $C_H$ ($C_H=C-C_K$) in the entire time window, shown in Fig. 1 as well. We see that the hydrodynamic effect of the Brownian particle is significantly larger when it has a bigger mass and a bigger size than a fluid molecule. At large times, the VACFs converge to the long-time tail of  $C(t)\sim t^{-3/2}$.

\begin{table}[!htbp]
\begin{tabular}{ccccccc}
\hline
$b(\phi)$ & 1(0.004) & 2(0.034) & 3(0.11) & 4(0.27) & 4.5(0.39) & 4.7(0.44)\\
\hline
$D_{CE}$ & 52.3 & 12.1 & 4.35 & 1.50 & 0.763 & 0.552 \\
\hline
$D_K$ & $52.3$ & $12.2$ & $4.35$ & $1.49$ & $0.760$ & $0.550$ \\
\hline
$D$ & 53.5 & 12.6 & 4.93 & 2.06 & 1.04 & 0.519 \\
\hline
\end{tabular}
\caption{Kinetic coefficients obtained by the Chapman-Enskog theory ($D_{CE}$) and by the decomposition ($D_K$). The diffusion coefficient $D$ is obtained by the Green-Kubo formula.}
\end{table}

{\bf Testing the Chapman-Enskog kinetic theory based on the decomposition.} The diffusion coefficient of identical hard spheres is predicted by the Chapman-Enskog theory in the first Sonine approximation~\cite{sl}, that is, $D_{CE}=\frac{3}{8nb^2 g(\rho)}(\frac{k_BT}{m\pi })^{1/2}$ with $g(\rho)=(1-\rho/2)/(1-\rho)^3$ and $\rho=4/3(n\pi b^3)$. In Tab. I, $D_{CE}$ obtained with this formula for different packing fraction is given, together with $D_K$ of identical hard spheres calculated by the decomposition method. Surprisingly, $D_K$ agrees with the theoretical prediction $D_{CE}$ perfectly over the entire fluid regime. However, in previous studies for testing the Chapman-Enskog theory~\cite{alder1970,alder2,sigurgeirsson2003transport}, $D_{CE}$ is compared with $D$, which may gives rise to ambiguous interpretation. To show this, in the Table we also provide $D$ calculated by the Green-Kubo formula with the integral time being truncated at $t=500$ (this truncation time is adopted throughout. The finite-size effect induced by such a truncation will be discussed later). We see that the deviation of $D$ from $D_{CE}$ may exceed $30\%$ in the high packing fraction region. Deviation of the same level have been encountered in previous studies~\cite{sigurgeirsson2003transport}. Calculating $D$ with a small $N$ may result in the decrease of the deviations and leads to $D$ be closer to $D_{CE}$~\cite{alder1970,alder2,sigurgeirsson2003transport}. Based on the comparison between $D$ and $D_{CE}$ one may conclude that the Chapman-Enskog theory is effective for small systems and in a short time period. But in contrast, the accurate agreement between $D_K$ and $D_{CE}$ suggests that, independent of the system size and the evolution time, $D_{CE}$ predicts $D_K$ accurately, implying that the Chapman-Enskog theory should be interpreted properly as a pure kinetic theory where the hydrodynamical contribution is not taken into account.

\begin{figure}[!htbp]
\centering
\includegraphics[width=\linewidth]{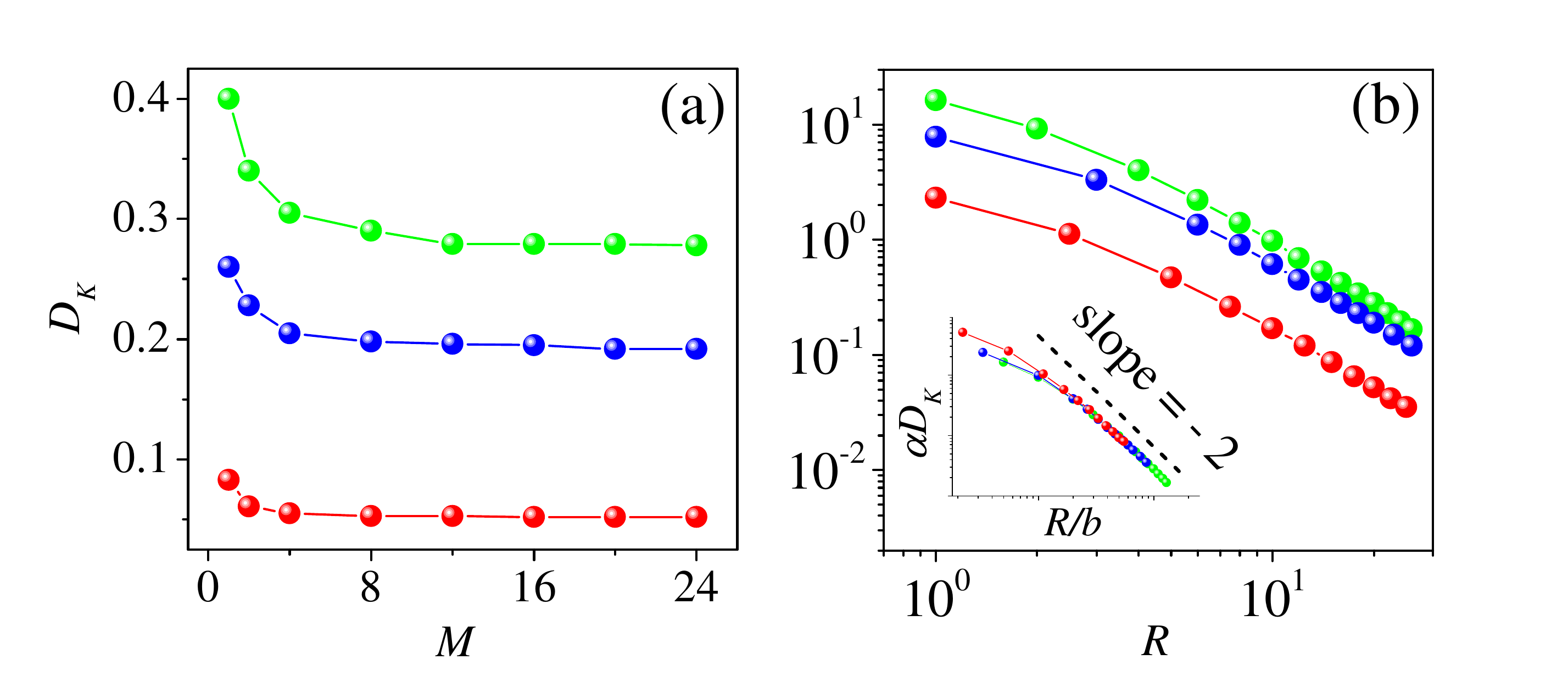}

{\caption{ $D_K$ of the Brownian particle as a function of $M$ at $R=20$ (a) and of $R$ at $M=24$ (b). Three curves (from top to bottom) are for $\phi=0.034, 0.11$, and $0.39$, respectively.}}
\label{zhqfig2}
\end{figure}

{\bf Examining the original and revised SE relations.} In Fig. 2 we show $D_K$ of the Brownian particle as a function of $M$ with $R=20$  [Fig. 2(a)] and as a function of $R$ with $M=24$ [Fig. 2(b)] for three packing fractions, $\phi=0.034(b=2)$, $0.11(b=3)$, and $0.39(b=4.5)$, respectively. Numerical errors, throughout this paper, are suppressed within the size of symbols by collecting a sufficient amount of ensemble samples. In Fig. 2(a), it shows that at a fixed $R$, $D_K$ converges to a mass-independent constant as $M$ increases, while in Fig. 2(b) it indicates that at a fixed mass, $D_K$ decreases as

\begin{equation}
{D_K }\sim R^{-2}
\end{equation}
when $R$ is large. Moreover, the behavior of $D_K$ versus $R$ can be re-scaled to fall on a universal function (i.e., multiplying it by a $b-$depenednt constant $\alpha$ and plotting it as a function of $R/b$, see the inset in Fig. 2(b)). This is a key evidence that the left-hand-side of the original SE relation Eq. (1) should depend on the Brownian particle's size, since it leads to $DR\sim c_1/R+D_HR$, where $c_1$ is a fitting parameter.

\begin{figure}[!htbp]
\centering
\includegraphics[width=\linewidth]{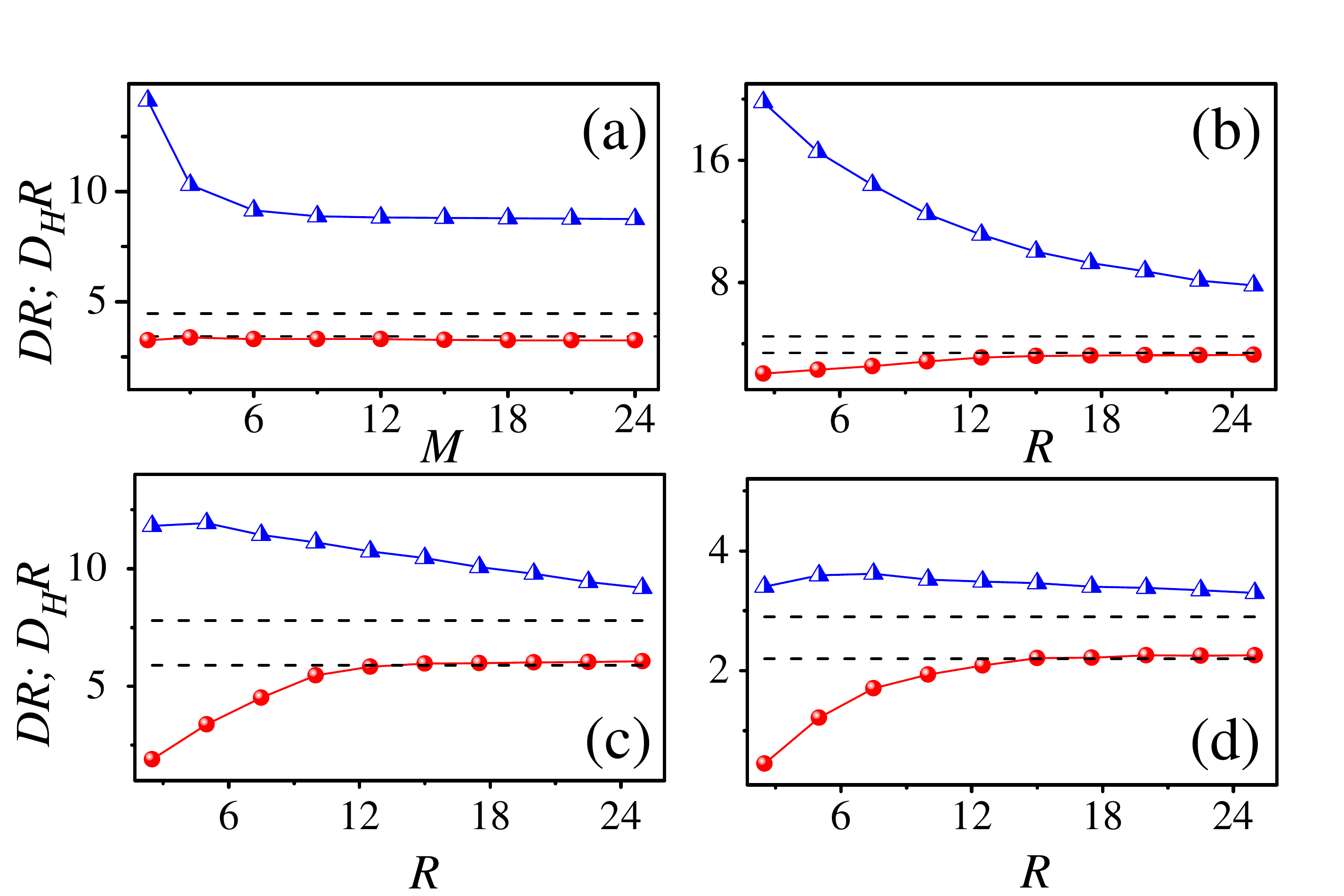}
{\caption{Examination of the SE relation. $DR$ (triangles) and $D_HR$ (circles) of the Brownian particle as a function of $M$ at $R = 20$ in the fluid of $\phi=0.034$ (a), and of $R$ at $M=24$ in the fluids of $\phi=0.034$ (b), $\phi=0.11$ (c), and $\phi=0.39$ (d), respectively. The top and the bottom dashed line in each panel indicates the predicated DR of Eq. (1), ${k_BT}/{c\eta}$, with $c=6\pi$ and $8\pi$, respectively.}}
\label{zhqfig3}
\end{figure}

We then examine the mass and size dependent behavior of $D_H$. Figure 3(a) reveals the $M$-dependence of $DR$ and $D_HR$ in the fluid of $\phi=0.034$. The size of the Brownian particle is fixed at $R=20$. We see that $DR$ varies at small masses and converges at large masses, while $D_HR$ keeps mass independent. Therefore, light Brownian particles do not obey the original SE relation. Since $D_H$ is mass independent, the mass-dependent behavior of $DR$ should be ascribed to the consequence of $D_K$.

Figure 3(b) shows the $R$-dependence in the fluid of $\phi=0.034$. The mass of the Brownian particle is fixed at $M=24$. We see that $DR$ decreases as $R$ increases, while $D_HR$ turns out to be size-independent for $R\ge15$. Note that the converged value of $D_HR$ in Fig. 3(b) is the same as that in Fig. 3(a). Therefore, $D_HR$ is neither mass nor size dependent for about $R>15$, indicating that the revised SE relation is applicable in this $R$ range. This fact also implies that $D_H \sim R^{-1}$ for large Brownian particles and the size dependence of $DR$ is caused by $D_K$.

Figure 3(c) and 3(d) show the $R$-dependence in higher packing fractions. The mass of the Brownian particle is fixed also at $M=24$. We see that they give the qualitatively similar results as in the Fig. 3(b). Note that usually $\phi=0.3$ is considered as the boundary between gas and liquid phases~\cite{ohtori2018stokes}. Thus, our studies here cover both the gas and liquid regime. Therefore, in the range of $R$ we have investigated, the original SE relation is obviously violated but the revised SE relation is respected for about $R\ge15$.

As $DR\sim c_1/R+D_HR$, $DR$ is expected to become approximately size independent at large enough $R$ when $c_1/R \ll D_HR$ and as a result, $DR\sim D_HR$. As shown in Fig. 3, at $R=15$, the deviation between $DR$ and $D_HR$ is about $300\%$, $170\%$, and $150\%$ for the three packing fractions. These differences should disappear eventually with the further increase of $R$. Fitting the data shown in Fig. 2(b) we obtain $D_K\sim c_1R^{-2}$ with $c_1\sim 110$, $80$, and $23$ for the three packing fractions. Then we can infer that when $R$ is larger than about $R\sim 2,500, 1,500,$ and $1,000$, respectively, the deviation between $DR$ and $D_HR$ will be smaller than $1\%$. As an estimation for real systems, we set the water molecule diameter (about $0.4$ nanometer) to be the unit length in our model, the remarkable size-dependent effect may disappear for Brownian particles up to a micron in gas, and to hundreds of nanometers in liquid. Note that, though keeping decrease continuous, the rate of decrease of $DR$ via $R$ should become much smaller in liquid than in gas since $c_1$ is much smaller in the former case, as Fig. 3 shown.

In Tab. I,  $D$ and $D_K$ for the particular case of $M=1$ and $R=b$ are listed. We find that, for example, in the fluid of $\phi=0.39$, we have $DR\sim 4.5$ and $D_HR\sim 1.3$, which are remarkably different from the results of large Brownian particles shown in Fig. 3. Therefore, neither the original nor the revised SE relations can stand the size-dependent test when the size of a Brownian particle decreases down to the molecular level.

The kinematic shear viscosity can be calculated following the method in Refs.~\cite{alder1970,pre06}, which gives $\nu=11.8\pm0.1,6.8\pm0.1$, and $18.3\pm0.2$ for the three packing fractions, respectively. Then, we can determine the parameter $c$ in the SE relation. In Fig. 3, the dashed lines represent $k_BT/c\eta$. Those with $c=6\pi$ corresponds to the slipping boundary conditions. We see that the converged $D_HR$ value does not agree with the predictions with $c=6\pi$. Instead, they are close to the predictions with $c=8\pi$.

The finite-size effects may contribute a remarkable correction to $DR$ and $D_HR$. When the Brownian particle is identical with the fluid molecules, there is a well-known empirical formula~\cite{yeh2004system}, $D(\infty)=D(L)(1-1.9b/L)^{-1}$, for estimating the diffusion coefficient $D(\infty)$ of infinite large system from that of a finite system, $D(L)$. It gives a $2\%$ lifting to $D(L)$ for fluid molecules of $b=3$, which is slight. There is no empirical formula for a general Brownian particle; in this case, we estimate the correction as following. Due to the computation complexity we have truncated the upper limit of integral of Eq. (3) at $t=500$. Within such an interval, we can verify that the system size of $N=27,000$ is large enough, since the VACF obtained with this size overlaps with that obtained with a much larger $N$ (see the Supplemental Materials). However, the truncated tail of the VACF may contribute a significant correction. For example, fitting the tail of $C(t)$ we obtain $C(t)\sim 0.6t^{-3/2}$ for a Brownian particle of $R=20$ and $M=24$ in the fluid of $\phi=0.11$, which may lift $DR$ and $D_HR$ about $9\%$ and $15\%$, respectively. The latter makes $D_HR$ close to the prediction with $c=6\pi$. Meanwhile, the finite-size effect of $\nu$ has not been considered. Due to the computation complexity we have to leave the accurate value of $c$ an open problem. For a more detailed discussion of the finite-size effects, see the Supplemental Materials.

{\bf The underlying mechanisms.} We propose a phenomenological understanding why the revised SE relation works well for large Brownian particles while becomes also size dependent for small ones. Our reasoning is similar to that for the long-time tail in Ref.~\cite{sl} (p. 246), but with the key difference that we take the delay effect into account. Suppose that at $t=0$, the Brownian particle resides at $\bf{r}=0$. Then, its initial momentum gradually transfers to the surrounding fluid molecules by collisions, and spreads out by the viscosity and sound modes. The momentum density $\Psi(r,t)$ carried by the viscosity mode relaxes diffusively as $\Psi(r,t) \sim \exp(-r^2 /(4\nu t))$. This mode feeds the momentum back to the Brownian particle and contributes to the hydrodynamic component $D_H$. The characteristic radius of the pack region of $\Psi(r,t)$ expands as $r_c=\sqrt{4\nu t}$. Assuming the Brownian particle floats at the average velocity of fluid molecules in this region, Eq. (5) is derived~\cite{sl}. Based on this reasoning, we further emphasize that for a Brownian particle of radius $R$, the viscosity mode is physically meaningless for $t<R^2/4\nu$ since $r_c<R$ in this time period. In other words, the feedback begins to play a role only after $t=R^2/4\nu$. Then, plugging Eq. (5) into Eq. (3), the lower limit of integral should be larger than $t=R^2/4\nu$. With this consideration we obtain an estimation of $D_H$, i.e., $D_H\approx \int_{R^2/4\nu}^\infty C_H(t)dt\sim \sqrt{\nu}/R(D_K+\nu)^{3/2}$. This result explains, on one hand, $D_HR\sim 1/\nu$ is size-independent for relatively large Brownian particles when $D_K$ can be neglected comparing with $\nu$. Extrapolating $D_K$ based on the scaling laws of $D_K \sim c_1R^{-2}$, we get that $R > 32, 34$, and $35 $ can assure $D_K/\nu < 1\%$ for the three packing fractions, respectively. These results agree in order with the simulation results of Fig. 3, where no remarkable deviations of $D_HR$ can be recognized for $R\ge15$. On the other hand, it indicates that $D_HR$ must be size dependent for small enough Brownian particles when $D_K\approx\nu$, where $D_K$ also contributes a $R$ dependent component. This feature implies that a full exclusion of the kinetic effect fails.

Though $D_HR$ may also be size dependent for small Brownian particles, the threshold of $R$ above which it becomes size independent is much smaller than that of $DR$. For $DR$, whether $D_K$ is negligible should be determined by comparison with $D_H$ (which is usually far smaller than $\nu$).

\section*{Summary and discussion}

Diffusion theories of a Brownian particle are usually established either in the kinetics or in the hydrodynamics framework, and therefore they should be applied on the basis of the decomposition. Excluding the hydrodynamic effect, we find that the Chapman-Enskog theory of the kinetic diffusion coefficient is perfectly accurate over the entire fluid regime. Excluding the kinetic contribution, we find that the SE relation can be extended to Brownian particles down to the size only several times larger than the fluid molecules; otherwise it depends significantly on the size of the Brownian particle till it is several hundred times the size of the fluid molecules, and depends obviously on the mass of the Brownian particle till it is over tenfold the mass of the fluid molecules. When the size of the Brownian particle is comparable with a fluid molecule, a full exclusion of the kinetic effect fails, and neither the original nor the revised SE relation is applicable. Efforts based on the structural entropy of fluids for finding a universal model different from the SE relation  ~\cite{rosenfeld1999quasi,dzugutov2001addendum,bretonnet2002self,kaur2005nature,ning2019universal} may provide a possible way to model the diffusion of fluid molecules.

We would like to point out that a fine experimental test of the size and mass dependence of the SE relation is becoming possible. Indeed, traditional experimental techniques~\cite{barhoum2016diffusion,mason1997particle,durr1998brownian,palit2017combining,lin2017determining,kim2017solutal} have already allowed one to measure the trajectory of a single Brownian particle of hundreds of nanometers. In recent years, thanks to the state-of-the-art experimental developments, the measurements of displacement as well as instant velocity of a single Brownian particle have been available~\cite{exp1,exp2,exp3,exp4,exp5,exp6,exp7}, and thus make it possible to obtain an accurate $C(t)$ experimentally and to decompose $D_K$ and $D_H$. We look forward to the experimental investigations in the near future.

\bibliography{brownianmotion}

\section*{Acknowledgements}
This work is supported by the NSFC (Grants No. 11335006).

\end{document}